# From Rights to Rites:
## Expectations Management in Smart-Home AI

Varad Vishwarupe[1,2]   Ivan Fléchais[1]   Marina Jirotka[1]   Nigel Shadbolt[1,2]

[1] *Department of Computer Science, University of Oxford*

[2] *Institute for Ethics in AI, University of Oxford*

{varad.vishwarupe, ivan.flechais, marina.jirotka, nigel.shadbolt}@cs.ox.ac.uk

**Abstract.** Domestic voice assistants and smart-home devices are increasingly embedded in everyday routines, yet their ethics are often treated as an afterthought or delegated to compliance teams. To explore how expectations about smart-home AI are constructed and managed, we conducted 33 semi-structured interviews with designers, developers, and researchers from major smart-home platforms (Amazon Alexa, Microsoft Azure IoT, and Google Nest). Using a constructivist grounded-theory approach, we develop *Expectations Management (EM)*: a culturally embedded model describing how practitioners shape, calibrate, and repair expectations by balancing organisational rights with culturally situated rites. We show that EM differs from expectation-confirmation theory and trust-calibration by foregrounding moral judgement, situated action, and cross-cultural variation. Our analysis reveals four recurring design tensions—automation vs. autonomy, helpfulness vs. intrusiveness, personalisation vs. predictability, and transparency vs. obscurity—and distils them into a five-phase EM Design Playbook that supports moral prudence. We discuss implications for responsible smart-home design and offer guidance for human-centred AI.

## 1  Introduction

Smart home assistants, thermostats, cameras, locks, and sensors are increasingly embedded in everyday domestic routines. What distinguishes contemporary smart homes from earlier forms of home automation is not merely connectivity, but the integration of AI-driven systems that infer preferences, predict behaviour, and act with limited human input.

While these systems promise convenience and efficiency, they also reshape how households coordinate, monitor, and interpret daily life. In contrast to many other computing environments, the home is a morally dense space: it is where norms around privacy, autonomy, authority, and cultural identity are continuously enacted and negotiated, often across unequal relationships of age, gender, and dependency.

As a result, many failures in domestic AI are not technical malfunctions but moments where systems behave as designed and still feel inappropriate, intrusive, or disrespectful. These misalignments often arise when design assumptions about authority, consent, visibility, or "normal" behaviour do not align with the social and cultural expectations of the household. In the home, such breakdowns are not experienced merely as usability problems; they are experienced as moral disruptions.

AI systems inevitably encode normative assumptions about how households ought to function. These assumptions become consequential not only through what systems do, but through what users expect them to do or to refrain from doing. Ethical friction thus emerges not only from error but from mismatches between encoded assumptions and culturally situated expectations. Managing expectations about what systems do, what they imply, and whose values they embody therefore becomes a central design challenge [1–4].

Classic expectation-confirmation theory (ECT) explains satisfaction as a comparison between pre-use expectations and post-use experience, treating expectations as cognitive standards formed prior to use [3, 4]. Trust-calibration frameworks similarly focus on aligning user trust with system capability to avoid overtrust and undertrust [6–8]. While these models have shaped research on intelligent systems, they offer limited guidance on how expectations are shaped through design decisions, how cultural norms influence what is perceived as appropriate behaviour, or how organisational policies and constraints are negotiated in practice [5, 7–9]. Moreover, they focus primarily on user attitudes after deployment rather than the upstream work of designers and developers who set the conditions under which expectations are formed.

Human–computer interaction research has long shown that interaction is situated and socially organised rather than fully determined by plans or specifications [1, 2]. In domestic settings, expectations about technology are embedded in household routines, relationships, and culturally situated notions of care, privacy, and respect [14, 15]. For smart-home AI, this means expectations are shaped upstream through design decisions such as defaults, tone, disclosure language, and autonomy bounds. In practice, expectations become most empirically visible when systems strain or violate these assumptions, prompting calibration or repair. We therefore treat breakdowns not as the origin of expectations, but as moments where implicit expectations surface and can be examined.

This paper studies how designers, developers, and researchers manage user expectations during the design and deployment of smart-home AI. By examining practitioner accounts, we capture how ethical and cultural expectations are anticipated, translated, tested, and repaired through concrete design practices, including onboarding flows, automation thresholds, policy-driven disclosures, and responses to user complaints or disengagement. This upstream perspective surfaces forms of ethical and interpretive labour that are difficult to explore from post-hoc user studies alone.

We address the following research questions:

**RQ1.** How do designers, developers, and researchers account for ethical and cultural expectations when designing smart-home AI?

**RQ2.** How do practitioners interpret and negotiate culturally situated domestic rites, including norms of care, privacy, hierarchy, and appropriate initiative?

**RQ3.** How are organisational rights—policy requirements, legal obligations, and safety standards—balanced against domestic rites in everyday design and deployment practice?

This paper makes three contributions. First, it develops a constructivist grounded theory of Expectations Management (EM) as a cyclical process of shaping, calibrating, and repairing expectations in domestic AI. Second, it identifies four recurring design tensions that structure this work: automation vs.

autonomy, helpfulness vs. intrusiveness, personalisation vs. predictability, and transparency vs. obscurity. Third, it distils practitioner strategies into a compact EM playbook framed as reflective guidance.

## 2   Background and Related Work

### 2.1   Smart-Home AI and Cultural Expectations

Smart-home ecosystems include voice assistants, connected lighting, thermostats, locks, cameras, and sensors that continuously collect data and increasingly act on inferred household routines and preferences. As these systems move beyond passive response toward proactive assistance, they enter domains shaped by cultural norms of care, privacy, hierarchy, and respect. Prior work shows that expectations surrounding assistance and intrusion vary significantly across cultural contexts [13, 14, 15]. In collectivist settings, proactive help may be interpreted as attentiveness or care, whereas in more individualistic contexts, similar behaviour may be experienced as invasive or inappropriate.

Empirical studies highlight these differences. Research on domestic voice assistants documents how families negotiate acceptable behaviour through social roles and household routines rather than formal settings alone [10, 11]. Large-scale surveys further indicate that voice assistants are perceived as particularly problematic with respect to privacy and security [13]. Together, these findings suggest expectation alignment in smart-home AI cannot be culturally universal and must account for situated domestic norms.

### 2.2   Expectation Confirmation and Trust Calibration

Expectation Confirmation Theory (ECT) models satisfaction as a cognitive comparison between pre-use expectations and post-use experience [3–5]. While influential, ECT treats expectations as relatively stable mental standards formed prior to interaction and offers limited insight into how expectations are shaped through design decisions or how they function as moral norms within domestic life.

Trust-calibration research focuses on aligning trust with a system's capabilities to avoid overtrust and undertrust [6–8]. These models emphasise transparency, feedback, and appropriate reliance. However, they often abstract away from cultural context and rarely examine how designers interpret, negotiate, and manage trust during development, or how repair functions as ethical restoration following breakdown rather than technical correction alone [9].

### 2.3   Situated and Practitioner-Centred Perspectives

HCI research argues that interaction is situated, emergent, and shaped by local context rather than fully determined by design-time plans [1, 2]. Domestic technology research further demonstrates that homes are moral and social spaces structured by routines, relationships, and power dynamics [14, 15]. For smart-home AI, this implies expectations are not simply held by users but enacted through ongoing interaction between households and systems.

Building on this tradition, we adopt a practitioner-centred perspective and examine Expectations Management (EM) as enacted by designers, developers, and researchers during the design and deployment of smart-home AI. Practitioners often first encounter expectation misalignment through complaints, disengagement, escalation, or post-incident review. These moments of breakdown make otherwise implicit expectations visible and actionable. By focusing on upstream work, our study complements user-centred research and foregrounds the ethical and cultural judgement exercised by practitioners as they balance organisational constraints with domestic norms.

## 3 Methods

### 3.1 Research Approach

We adopted a constructivist grounded-theory (CGT) approach [18] to develop an empirically grounded account of how designers, developers, and researchers manage user expectations in smart-home AI. CGT is well-suited here because it foregrounds meaning-making, reflexivity, and theory-building from practice accounts rather than hypothesis testing. We model Expectations Management (EM) as a process enacted through design and deployment decisions under conditions of ethical ambiguity, cultural variation, and organisational constraint.

### 3.2 Participants and Recruitment

We conducted 33 semi-structured interviews with practitioners working in smart-home AI divisions at Amazon (Alexa), Google (Nest), and Microsoft (Azure IoT). Participants were sampled across roles because expectations management is distributed across design, engineering, research, and product functions. Participants were recruited through professional networks and snowball sampling to capture variation in role, seniority, and organisational context. Interviews were conducted between February and April 2025, primarily in person at corporate campuses in India and the UK, with a small number conducted online. Interviews lasted 45–90 minutes (mean 65 minutes), yielding approximately 35.8 hours of audio. To reduce identifiability, participants are referred to as P1–P33, with identifiers assigned randomly and used uniformly throughout. Table 1 summarises participant characteristics.

**Table 1.** *Participant overview.*

| Dimension | Summary |
| --- | --- |
| **Roles** | Designers and UX researchers (19); developers and applied AI engineers (10); product managers and research roles (4) |
| **Experience** | 2 to 15+ years |
| **Product scope** | Voice assistants; smart speakers and displays; routines and automations; home security devices; IoT platform infrastructure |
| **Market exposure** | Primarily India- and UK-focused products and localisation work, with some involvement in global launches |
| **Organisations** | Amazon Alexa; Google Nest; Microsoft Azure IoT |

### 3.3 Data Collection

Interviews were semi-structured and focused on: (1) how practitioners interpret user expectations and cultural variation; (2) how organisational rights (policy, safety, legal obligations) shape design decisions;

and (3) how teams respond to expectation violations. Participants were asked to describe concrete episodes such as feature rollouts, localisation decisions, support escalations, post-incident reviews, and situations that felt ethically uncomfortable. Reflexive notes and analytic memos were written immediately after each interview.

### 3.4 Analysis

Analysis followed CGT procedures using open, focused, and theoretical coding [18]. Line-by-line open coding was conducted by the first author to maintain analytic closeness to participant language and preserve cultural nuance. Focused coding clustered recurring codes into categories via constant comparison across roles and organisations. Theoretical coding integrated relationships among categories into the EM cycle of shaping, calibrating, and repairing expectations, as well as the tensions that cut across these stages. Regular peer debriefs within a human-centred AI research group were used to challenge interpretations and reduce single-analyst drift. Table 2 illustrates the coding progression.

Table 2. *Illustrative coding progression from transcript fragment to EM stage.*

| Transcript fragment | Open code | Focused category | EM stage |
|---|---|---|---|
| *"We follow GDPR but soften the tone; ethics should still feel polite." (P10)* | Polite compliance | Rights translated into interaction | Shaping |
| *"Sometimes extended silence says more than feedback forms." (P11)* | Silence as boundary signal | Reading moral comfort cues | Calibrating |
| *"When the product says 'sorry', it's a cultural gesture." (P15)* | Apology as cultural act | Ethical restoration | Repairing |

### 3.5 Ethics and Positionality

The study received ethics approval from the Department of Computer Science Research Ethics Committee at the University of Oxford (CUREC ID 210291). All participants provided informed consent, and data were anonymised prior to analysis. The first author's position as an Indian researcher trained in Anglo-European ethics traditions provided insider sensitivity to culturally specific cues while also risking imported normative assumptions; reflexive memoing was used throughout to surface and contest these assumptions. While the data capture expert perspectives rather than end-user experience directly, this focus is intentional and aligned with the aim to examine upstream expectations management during design and deployment.

## 4 Findings

### 4.1 Overview: Grounded Theory of Expectations Management

We define Expectations Management (EM) as follows:

> *Expectations Management (EM) is a cyclical, culturally embedded form of design work through which designers, developers, and researchers manage users' expectations while developing and deploying smart-home AI systems by balancing organisational* rights—*policy, legal requirements, safety standards, and compliance obligations—with users'* rites, *understood as culturally situated norms of care, privacy, hierarchy, and respect.*

Expectations Management did not emerge as a linear design stage. Instead, it is recurring interactional work. Expectations are shaped through upstream design decisions (defaults, tone, autonomy bounds, disclosure practices), recalibrated as systems encounter everyday domestic use, and repaired when system behaviour violates what households take to be appropriate.

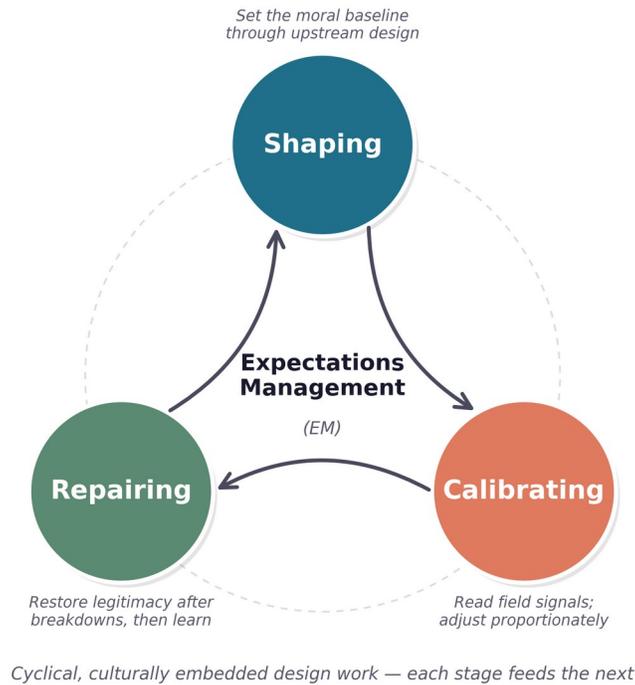

*Cyclical, culturally embedded design work — each stage feeds the next*

**Fig. 1.** *The Expectations Management lifecycle—shaping, calibrating, and repairing—with feedback loops across stages.*

## 4.2 Shaping Expectations

Shaping is the pre-deployment phase where practitioners establish the ethical and affective baseline of a system. Participants described three key inputs.

**Cultural and relational values.** Norms of care, politeness, and shared privacy inform what users consider respectful behaviour. Practitioners described how culturally situated expectations influence tone, deference, and acceptable initiative.

**Organisational influences.** Brand narratives (e.g., assistant-as-companion) and default policies communicate role boundaries and permissible conduct.

**Professional and personal values.** Practitioners draw on human-centred design, accessibility, and their own moral sensibilities when translating rights into interaction.

Through shaping, teams embed foresight into defaults, onboarding flows, and disclosure language. For example, one participant explained how recording announcements required by policy were softened to avoid interrupting conversation: "We follow GDPR but soften the tone; ethics should still feel polite" (P10).

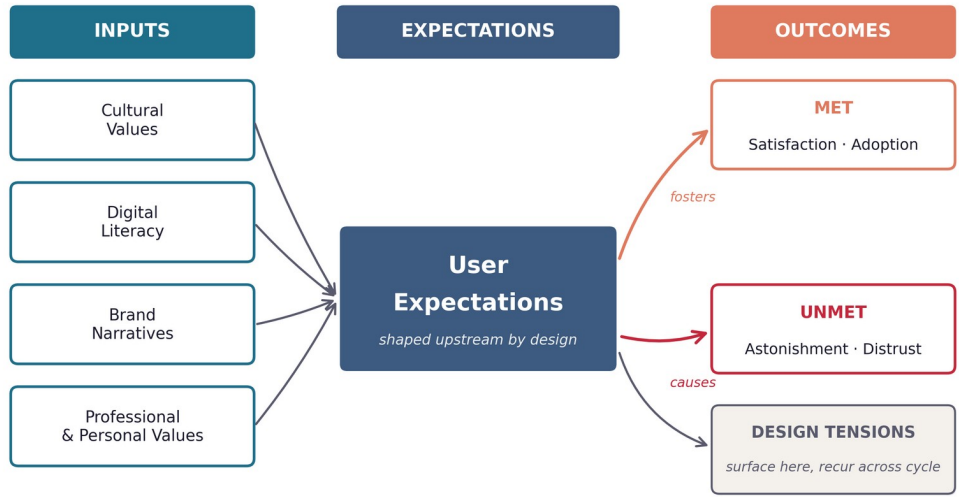

**Fig. 2.** *Shaping expectations: inputs from cultural, organisational, and professional values feed user expectations, which yield met outcomes (satisfaction, adoption) or unmet outcomes (astonishment, distrust) that, in turn, surface design tensions.*

## 4.3 Calibrating Expectations

Calibration is the ongoing work of keeping behaviour proportionate as contexts shift. Unlike shaping, calibration is driven by field evidence rather than design intent alone. Participants described three recurring practices.

First, practitioners learned to read moral signals from interaction—including hesitation, disengagement, or extended silence—treating these as indicators of discomfort rather than mere usability friction. Second, teams distinguished ordinary interactional noise from culturally meaningful violations, which shaped whether responses were treated as tuning issues or ethical concerns requiring redesign. Third, calibration involved proportional adjustment: modifying autonomy thresholds, timing, disclosure levels, or rollout strategies to restore balance rather than removing features outright.

Calibration extends trust calibration, but the locus differs: the central concern was not only whether trust matched capability, but whether the system exercised restraint in culturally legible ways, especially where rights under-specified what respectful enactment should
*feel* like at home.

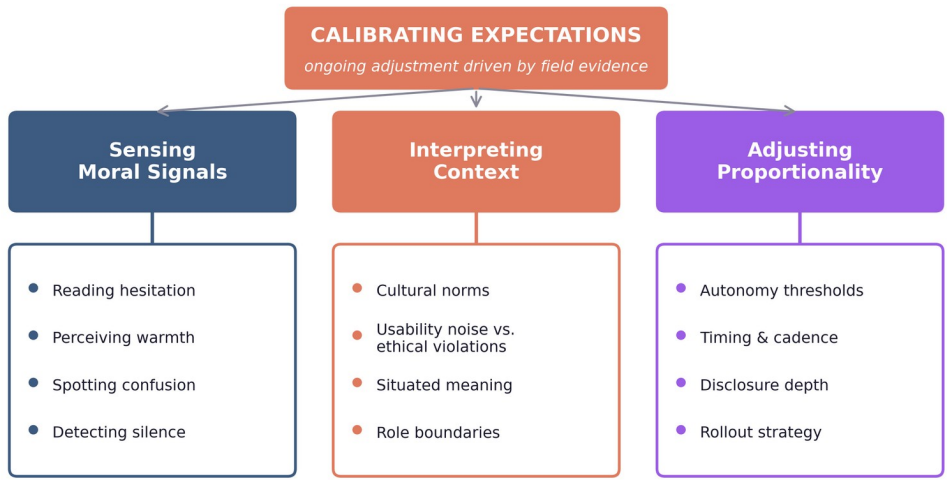

**Fig. 3.** *Calibrating expectations: continuous adjustment based on field signals and practitioner judgement, organised around sensing, interpretation, and proportional response.*

## 4.4 Repairing Expectations

Repair occurs when expectations are violated and legitimacy is disrupted. Participants consistently distinguished technical correction (functional faults) from ethical restoration (acknowledging social meaning and re-establishing respect). Repair was structured around three pillars.

**Accountability:** acknowledging what occurred and clarifying responsibility.

**Empathy:** tone and transparency recognising the emotional dimension of disruption.

**Continuity:** feeding incident learnings back into shaping and calibration.

Repair completes the cycle. Breakdown exposes implicit expectations, repair clarifies the boundary, and learning reshapes subsequent design decisions. In this way, repair is not merely a failure state but a mechanism for refining understandings of cultural and moral expectations in domestic settings.

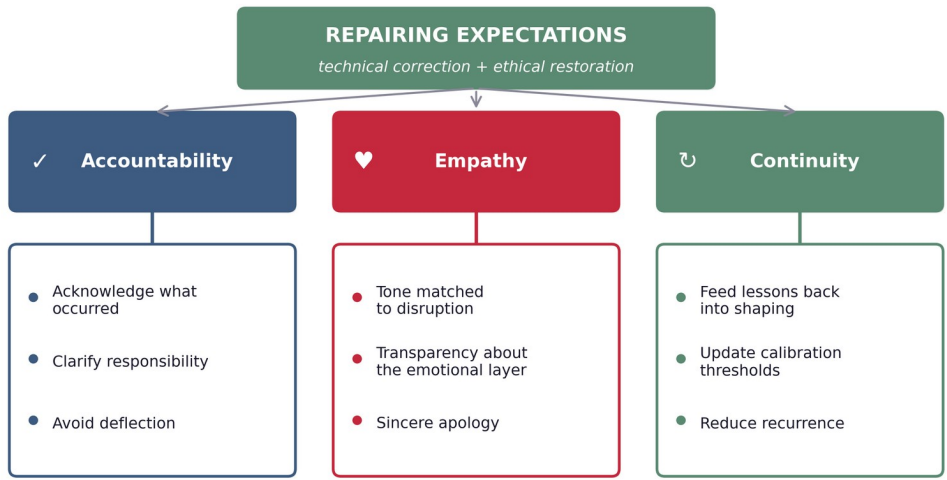

**Fig. 4.** *Repairing expectations: technical correction and ethical restoration as complementary work, organised around accountability, empathy, and continuity.*

## 4.5  Design Tensions

Across shaping, calibration, and repair, four recurring tensions structured Expectations Management. These tensions are not problems to be resolved once; they are conditions to be managed through situated judgement. Table 3 summarises each tension alongside its recurring practitioner strategy.

**Table 3.** *Four recurring tensions in EM and the practitioner strategies that manage them.*

| Tension | What is at stake | Recurring strategy |
| --- | --- | --- |
| **Automation ↔ Autonomy** | Convenience can erode the household's sense of agency. | Bounded delegation, reversibility, explicit permissioning. |
| **Helpfulness ↔ Intrusiveness** | Identical behaviours read as care or interference depending on context. | Tone, timing, and refusal pathways calibrated to preserve dignity. |
| **Personalisation ↔ Predictability** | Adaptive learning can feel like surveillance. | Staged consent and limits on learning in sensitive domains. |
| **Transparency ↔ Obscurity** | Full disclosure can overwhelm; silence can feel deceptive. | Minimal persistent cues paired with on-demand explanation. |

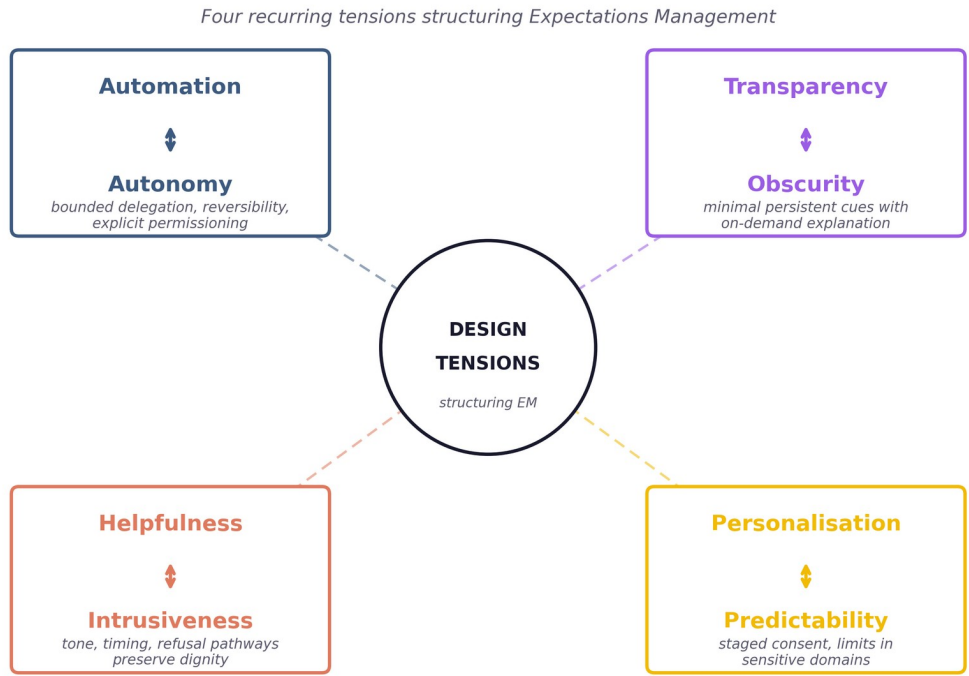

**Fig. 5.** *Design tensions that structure Expectations Management across lifecycle stages.*

## 5 Discussion

### 5.1 Beyond Expectation-Confirmation and Trust Calibration

Expectations Management intersects with expectation-confirmation and trust calibration in recognising that expectations and trust shape adoption, reliance, and satisfaction [3, 4, 6]. The difference lies in analytic emphasis. ECT treats expectations primarily as user-side reference points compared against outcomes. EM shows how expectations are actively produced upstream through design decisions, including defaults, tone, autonomy bounds, and disclosure practices. Expectations are not only held by users; they are built into interactional form.

Trust-calibration frameworks align trust with technical capability [6–8]. EM extends this by showing that capability alone is insufficient in domestic contexts: systems may function correctly and still be rejected if they violate norms of restraint, politeness, or care. EM foregrounds calibration of interactional propriety, not only reliability. Finally, EM treats repair as constitutive rather than secondary. In domestic AI, legitimacy is often restored through acknowledgement, explanation, and redesign rather than technical fixes. This positions repair as a core site of ethical work.

### 5.2 Design and Development as Sites of Expectations Work

Our focus on designers, developers, and researchers is deliberate. They translate organisational rights into interaction—how compliance is voiced, how autonomy is gated, what is enabled by default, and what form an apology takes—and encounter breakdowns at scale through development-side channels (support escalations, incident reviews, telemetry, user research). Studying these practitioners reveals upstream moral and cultural reasoning that shapes what users later encounter at home.

## 5.3 Expectations Management as Practical Judgement

Across accounts, participants repeatedly described situations where policy and standards did not uniquely determine what action was appropriate, and where cultural variation made universal rules brittle. In such moments, expectations management required judgement rather than rule-following. This aligns with Aristotelian phronēsis (practical wisdom): choosing proportionate action in context where rules under-specify what respect requires. In EM, phronēsis appears most clearly in calibration and repair, where teams interpret signals, infer meaning, and respond with restraint, clarity, and sincerity.

## 5.4 Expectations Management as Reflective Guidance

Rather than a prescriptive checklist, we present the EM Design Playbook as a compact reflective tool derived from recurring strategies observed across the data (Fig. 6; Table 4). The playbook makes expectations management actionable without assuming that ethical or cultural judgement can be reduced to rules.

**Table 4.** *The five-phase EM Design Playbook with practitioner illustrations.*

| Phase | Focus | Practitioner illustration |
|---|---|---|
| Calibrate | Cultural & ethical grounding | *"The rule is fixed, but how it shows up at home is a design decision. If that first interaction feels rude, you lose trust before the product even starts."* (P4) |
| Conserve | Protect the moral baseline | *"Nothing broke technically, but over releases the assistant started feeling pushy. Conserving the original restraint is harder than adding features."* (P12) |
| Challenge | Test new autonomy gradually | *"We learned to treat autonomy like a conversation, not a switch. You don't jump levels without letting people say no."* (P19) |
| Communicate | Make behaviour legible | *"If people always know when the system is active, you don't need to explain everything all the time."* (P7) |
| Correct | Repair as technical + moral act | *"Fixing the bug is easy. Admitting we crossed a line, and changing the behaviour so it doesn't happen again, that's the real work."* (P23) |

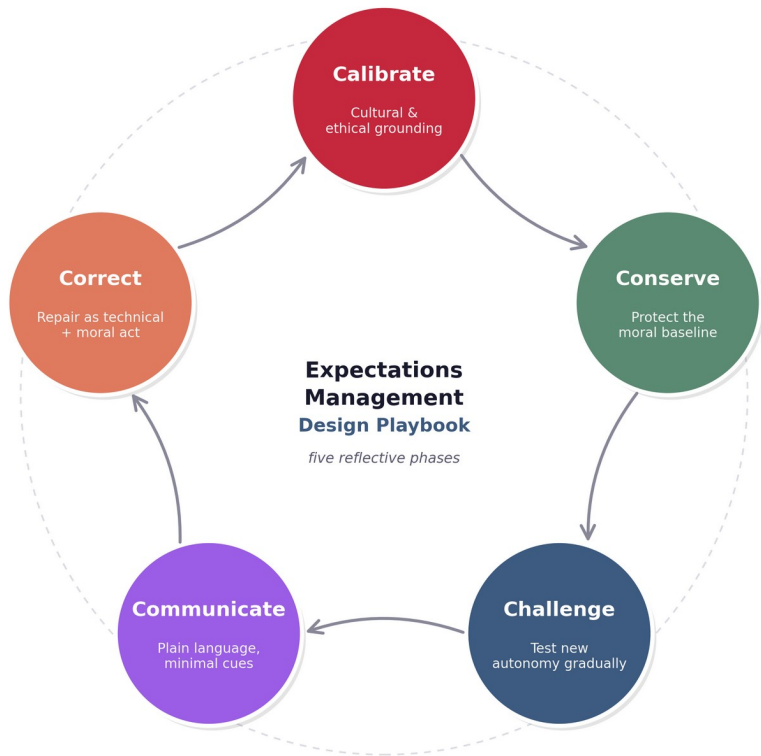

**Fig. 6.** *The EM Design Playbook: five reflective phases derived from recurring practitioner strategies.*

## 6  Scope, Limitations, and Future Directions

This study examines expectations management through interviews with practitioners in large smart-home platforms with mature policy, safety, and compliance regimes; findings may not transfer directly to smaller firms or informal deployments where such structures are weaker. We analyse practitioner accounts as situated narratives of ethical reasoning in design, rather than as direct representations of household experience. This is a deliberate choice: expectations are treated as practical commitments that surface through breakdowns and repair work encountered by teams (support escalations, incident reviews, telemetry patterns, user research artefacts). Future work should triangulate EM with household studies, including longitudinal and cross-cultural observation of calibration and repair as experienced by users.

Finally, expectations management unfolds within organisational constraints (release cycles, legal risk, business priorities). Further research could examine how incentives, tooling, and governance structures support or constrain EM practices, and how reflective judgement is sustained or eroded under production pressure.

## 7  Conclusion

This paper reframes ethical alignment in smart-home AI as ongoing socio-technical work rather than a one-time configuration problem. Based on interviews with 33 designers, developers, and researchers across Amazon Alexa, Google Nest, and Microsoft Azure IoT, we develop Expectations Management as a constructivist grounded theory of how user expectations are shaped, calibrated, and repaired by balancing organisational rights with users' rites.

Expectations surfaced most clearly through breakdowns and repair episodes, where practitioners interpreted signals of what counted as respectful, intrusive, or trustworthy conduct. By foregrounding these moments, EM shows how culturally and ethically aligned behaviour is sustained through restraint, translation, and sincere repair rather than technical performance alone.

The EM model and accompanying playbook contribute both a theoretical lens and reflective guidance for designing AI systems in intimate domestic spaces, where legitimacy depends not only on what systems can do, but on how they behave when expectations are tested. As smart-home AI continues to integrate into intimate spaces, expectations management will become increasingly important. Designers must move beyond static checklists and embrace moral prudence: anticipating, sensing, and responding to users' cultural expectations while upholding organisational rights in ways that support trust, legitimacy, and human flourishing.